\documentclass[letterpaper,dvipsnames]{article}
\pdfoutput=1
\usepackage{jheppub}
\usepackage{amsmath}
\usepackage{graphicx}% Include figure files

\usepackage{array}
\usepackage{xcolor}
\usepackage[normalem]{ulem}
\newcommand{\stkout}[1]{\ifmmode\text{\sout{\ensuremath{#1}}}\else\sout{#1}\fi}

\usepackage{slashed}
\usepackage{afterpage}
\usepackage{appendix}
\usepackage{multirow}
\usepackage{float}
\usepackage{mathtools}
\mathtoolsset{centercolon}

\usepackage{multirow}

\newcommand{\B}{\mathcal{B}}

\allowdisplaybreaks

\newcommand{\AddrOXF}{%
Rudolf Peierls Centre for Theoretical Physics, University of Oxford, Parks Road, Oxford OX1 3PU, UK
}

\newcommand{\AddrJHU}{%
Department of Physics and Astronomy, Johns Hopkins University, Baltimore, MD 21218, USA
}

\newcommand{\AddrCALTECH}{%
Division of Physics, Mathematics, and Astronomy, California Institute of Technology, Pasadena, CA 91125, USA
}

\title{Searching for axion forces with spin precession in atoms and molecules}

\author{Prateek Agrawal,}
\affiliation{\AddrOXF}

\author{Nicholas R. Hutzler,}
\affiliation{\AddrCALTECH}

\author{David E. Kaplan,}
\affiliation{\AddrJHU}

\author{Surjeet Rajendran,}

\author{and Mario Reig}
\emailAdd{mario.reiglopez@physics.ox.ac.uk}

\abstract{We propose to use atoms and molecules as quantum sensors of axion-mediated monopole-dipole forces. We show that electron spin precession experiments using atomic and molecular beams are well-suited for axion searches thanks to the presence of co-magnetometer states and single-shot temporal resolution. Experimental strategies to detect axion gradients from localised sources and the earth are presented, taking ACME III as a prototype example. Other possibilities including atomic beams, and laser-cooled atoms and molecules are discussed.}

\begin{document}

\maketitle
\section{Introduction}
Axions are well-motivated pseudo-scalar particles beyond the Standard Model (SM). Due to their appearance in the Peccei-Quinn solution to the strong CP problem \cite{Weinberg:1977ma,Wilczek:1977pj}, their role as dark matter \cite{Abbott:1982af,Preskill:1982cy,Dine:1982ah} and ubiquity in String Theory compactifications \cite{Svrcek:2006yi,Arvanitaki:2009fg} they have been receiving increased attention recently in both theory and experiment. 

On the experimental side, this surge in interest has led to a variety of searches both for cosmological relic axions, which contribute to the dark matter (DM) abundance, and DM-independent searches in the lab. Due to their CP-conserving dipole coupling to fermions, $c_\psi\frac{\partial_\mu \phi}{f_\phi}\Bar{\psi}\gamma^\mu\gamma^5\psi$, spin precession experiments are particularly appealing to look for these particles. Indeed, in the non-relativistic limit a coherent axion field $\phi$ gives rise to an energy shift with a spin $\mathbf{S}$ given by the interaction Hamiltonian \cite{Graham:2017ivz}:
\begin{equation}\label{eq:monopole-dipole_Hamiltonian}
    H_{\phi}=-\frac{1}{f_a}\mathbf{\nabla} \phi \cdot \mathbf{S}\,.
\end{equation}
Analogously to the well-known electromagnetic (EM) effects, in the presence of an axion background, the spin will precess around the gradient $\mathbf{\nabla} \phi$. The origin of this gradient can be either a relic axion DM background, or a sourced coherent axion  field. 

The last possibility is particularly interesting in the case where the axion has a Yukawa-like scalar coupling to nucleons, $g_s\phi \bar{N}N$. In this scenario the axion mediates a new kind of long-range interaction known as monopole-dipole force \cite{Moody:1984ba}, usually given in terms of the potential:
\begin{align}\label{eq:potential}
    V(r)
    &=
    \frac{g_s g^\psi_p}{8\pi m_\psi}
    \left(
    \frac{1}{\lambda_\phi r}+\frac{1}{r^2}
    \right)
    e^{-m_\phi r}\mathbf{S}\cdot \hat{r}\,,
\end{align}
with $m_\phi$ the mass of the axion,  $\lambda_\phi\sim m_\phi^{-1}$ the associated wavelength setting the effective reach of the force\footnote{See \cite{Banks:2020gpu} for a discussion about generalised potentials and their phenomenology in low-energy experiments.}, and $m_\psi$ the fermion mass. The couplings $g_s$ and $g_p=c_\psi\frac{m_\psi}{f_\phi}$ are the so-called monopole (CP violating) and dipole (CP preserving) coupling, respectively. Despite the expectation that these couplings should be very small to satisfy existing bounds, the coherent effect of around an Avogadro's number of source particles builds up leading to a potentially observable, macroscopic effect on a detector made of polarised spins.

The interaction of an electron with a combined magnetic $\mathbf{B}$ and axion $\phi$ field is given by: 
\begin{eqnarray}
    H & = & -g_e\mu_B\mathbf{S}\cdot\mathbf{B} -\frac{1}{f_a}\mathbf{S} \cdot (\mathbf{\nabla} \phi) \nonumber \\
    & = & -\mathbf{S}\cdot\left( g_e\mu_B\mathbf{B} + \mathbf{\nabla}\phi/f_a \right),\label{eq:spin_eq}
\end{eqnarray}
where $g_e$ is the electron $g$-factor and $\mu_B$ is the Bohr magneton.  The gradient of the axion field therefore acts similarly to a magnetic field in that it causes electron spin precession.
 Axion mediated forces can therefore be searched for with precision electron spin precession experiments, similar to those used in electron electric dipole moment (EDM) searches or in precision magnetometry, though with some important differences which we will discuss.

The current experimental bounds on axion-mediated forces on electron spins are given in Fig.\ref{fig:axion_electron_forces}. 
In this Letter we propose to use atomic and molecular beams and traps to look for axion-mediated forces, showing that these types of experiments have a promising potential for axion searches. To this end, we study the expected reach and describe qualitatively the main systematic effects and how to control them. As a specific example, we consider how ACME III could be adapted to search for these new macroscopic axion forces. We also propose a dedicated axion gradient-specific experiment using a beam of ytterbium $^{171}$Yb, and discuss the possibility of using laser-cooled molecules for axion force searches.   Note that using atomic and molecular EDM searches to put limits on axionlike particles has been previously considered in other contexts, for example via couplings within the atoms or molecules~\cite{Stadnik:2017hpa,Maison2021ALP,Prosnyak2023HfFAxion}, or via oscillating EDMs~\cite{Budker:2013hfa,Roussy2021HfFALP}.

\subsection{Geometry of an axion search with a beam experiment}
Spin precession experiments using the Ramsey method of separated oscillating fields constitute one of the most efficient methods to measure magnetic and electric dipole moments. By creating a superposition state and measuring the relative phase of the eigenstates after some time, very small energy splittings can be measured. This phase builds up during the spin coherence time, $\tau$, and the sensitivity of the experiment is proportional to $\propto\tau\sqrt{N}$, where $N$ is the number of particles measured. 

Experiments searching for the electron EDM $(\textbf{d}_e)$, for example, are designed to measure small energy shifts of the type $H_{\text{edm}}=-\textbf{d}_e\cdot \textbf{E}$, and can in principle also measure energy splittings from an axion field provided the geometry of the experiment is appropriate. 
As we have seen in Eq.(\ref{eq:monopole-dipole_Hamiltonian}) an  axion field  generates an energy splitting which depends on the relative orientation between the spin and the gradient.
The axion contribution to the phase will only be constructive if the orientation of the gradient and the quantisation axis is maintained during the coherence time\footnote{For example, this requirement is not satisfied in some EDM searches with ion traps \cite{PhysRevLett.119.153001,Roussy:2022cmp}, leading to an axion gradient effect which averages out. This kind of search will be sensitive only to spatial variations of the axion gradient and not to the gradient itself since a molecule in an ion trap has a rotating quantisation axis set by a rotating electric field. We thank Kia Boon Ng for pointing out this.}. 

To achieve their full sensitivity, experiments using spin precession in atoms or molecules usually require that the configuration of the experiment, such as the direction of the relevant fields, can be switched to measure differences in the spin precession frequency. This makes the experiment much more robust against slow drifts and offsets, which can be challenging to overcome. In the case of a coherent axion field sourced by a test mass, in addition to aligning the gradient along the quantisation axis, the ability to oscillate or reverse the position of the mass within periods of $\sim O(1)$ seconds, or faster, is very helpful. The effect induced by the axion gradient on the spins will be obtained from the change to the precession frequency that is correlated with the position of the mass. The distance from the axion field source to the spins sets the smallest (largest) wavelength (mass) that can be tested. This method has been used by SMILE \cite{Lee:2018vaq} and QUAX \cite{Crescini:2017uxs,Crescini:2020ykp}, setting the strongest lab bounds in the short range regime.A similar scheme will be employed in ARIADNE to test the monopole-dipole interaction on nucleon spins \cite{Arvanitaki:2014dfa}.

We now turn to the question of measuring an axion gradient from the earth's nucleons. This radial field will induce a DC signal on spins that cannot be reversed and therefore it seems difficult to measure it reliably; in particular, eq. (\ref{eq:spin_eq}) suggests that the axion gradient is indistinguishable from an uncontrolled background magnetic field, which will always be present\footnote{As shown in \cite{Agrawal:2022wjm,Davoudiasl:2022gdg}, storage rings are well-suited to search for this kind of DC axion signals. In the context of axion forces on electrons, the proposals in \cite{Suleiman:2021whz,Brandenstein:2022eif} are particularly promising. In this paper, however, we will concentrate on the opportunities at experiments using atoms and molecules. 
}. However, as we will show later, experiments using molecular or atomic beams provide an opportunity to measure this DC signal thanks to the presence of co-magnetometer states and single-shot temporal resolution   which offer robust methods to measure small, absolute spin precession signals not correlated with a magnetic field.

\section{Axion experiments with atoms and molecules}
Beam experiments using atoms and molecules have several features that make them particularly interesting to look for axion forces. 
These experiments usually have good sensitivity to quasi-DC signals, that is, they are well-suited to observe differences in the frequency as the experiment conditions are changed within a $\sim O(1)$ second scale. 
This is convenient, for example, when the source masses are relatively heavy, and cannot be moved to frequencies higher than $O(10)$ Hz.

In this section we first discuss different possibilities for measuring axion gradients with molecule and atom beam experiments. 

\begin{figure*}[t]
	\centering
	\includegraphics[width=0.75\textwidth]{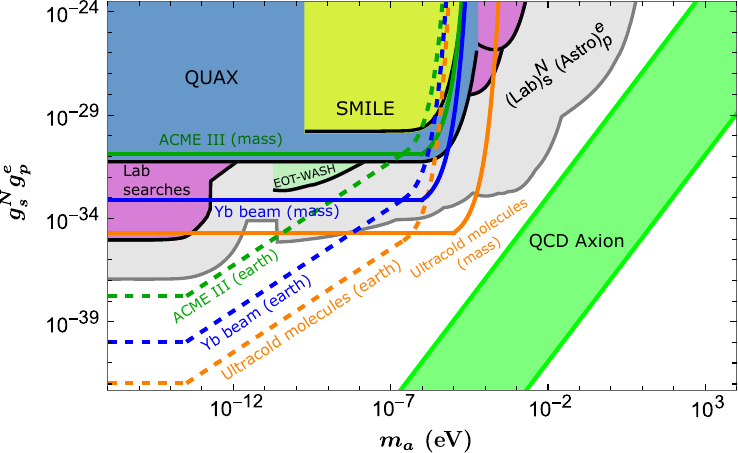}
		\caption{Axion-mediated monopole-dipole forces on electrons at spin precession experiments. The QCD axion prediction is shown in light green, taking $\theta_{\text{eff}}$ to lie in the range $10^{-20}<\theta_{\text{eff}}<10^{-10}$. The green solid (dashed) line corresponds to the expected sensitivity at ACME with moving test masses (earth) acting as the source (assuming that systematic effects and background are kept under control). The blue lines stand for the (shot noise-limited) sensitivity estimates a dedicated spin precession experiment using Yb beams. Finally, in orange, we have the expected reach with an experiment using ultracold molecules, assuming the existence of co-magnetometer states (assuming control of background and systematic effects). In either case, and specially thanks to the ability to test the earth gradient, new parameter space beyond astrophysical bounds will be covered. Bounds adapted from \cite{OHare:2020wah}.}

	\label{fig:axion_electron_forces}
\end{figure*}

\subsection{The need for co-magnetometry}
\label{sec:axion_comagnetometry}
As mentioned earlier, and shown in Eq. (\ref{eq:spin_eq}), an axion gradient looks very similar to a magnetic field pointing in the direction of the axion gradient.  In principle one could search for the axion field by performing a spin precession measurement in zero magnetic field, but this presents practical limitations.  In particular, there will always be some magnetic field component along every direction; every real material is slightly magnetic, and real magnetic shields must have holes in them for experimental access.  This problem is overcome in some EDM experiments by employing co-magnetometer schemes: the use of a species (or other internal state) with different relative magnetic and EDM sensitivity.  The Tl EDM experiment~\cite{Regan2002TlEDM} used a co-propagating beam of Na atoms, which are effectively insensitive to the eEDM, as an indepent measurement of the magnetic field.  The ACME~\cite{ACME:2018yjb} and JILA~\cite{Roussy:2022cmp} experiments use pairs of internal molecular states where the relative orientation of the internuclear axis and electron spin are different, thereby giving similar magnetic but opposite EDM sensitivity.  This ``internal co-magnetometry'' scheme is very powerful, but will unfortunately not work for the problem at hand.

To understand why, consider eq. (\ref{eq:spin_eq}); the electron spin interacts with the sum of the magnetic and axion terms, and therefore cannot distinguish between them.  However, this difficulty can be circumvented if the species has orbital angular momenta which provides a Zeeman interaction but does not couple to the axion field (as it is not a spin).  For example, consider an atom with electron spin $\mathbf{S}$, electron orbital angular momentum $\mathbf{L}$, and spin-orbit coupling $\beta$, so that the Hamiltonian for an atom interacting with a magnetic and axion field is given by
\begin{equation}
H = -\mu_B(\mathbf{L} + 2\mathbf{S})\cdot\mathbf{B}+\beta\mathbf{L}\cdot\mathbf{S} - \mathbf{S}\cdot\mathbf{\nabla}\phi/f_a,
\end{equation}
where we have set the electron $g-$factor to be 2.  In the physically relevant limit where $\beta$ is much larger than any other energy scale in the problem, the good quantum number is $\mathbf{J}=\mathbf{S}+\mathbf{L}$, and the energy shifts in a magnetic field are given by
\begin{equation}
    \Delta E_B = -g_J M_J\mu_B |\mathbf{B}|,
\end{equation}
where
\begin{equation}
    g_J = 1 + \frac{J(J+1) + S(S+1) - L(L+1)}{2J(J+1)}
\end{equation}
is the Land\'{e} $g$-factor and $M_J$ is the projection of $\mathbf{J}$ on the quantization axis.  We can go through a similar argument used to derive this equation to find the energy shift from an axion gradient, 
\begin{eqnarray}
      \Delta E_\phi & = & -\left\langle \mathbf{S}\cdot \mathbf{\nabla}\phi/f_a  \right\rangle \\
      & = & -\left\langle \mathbf{J}\cdot \mathbf{\nabla}\phi/f_a  \right\rangle \left\langle \frac{\mathbf{S}\cdot\mathbf{J}}{|\mathbf{J}|^2} \right\rangle \\
      & = & -g_a M_J |\mathbf{\nabla}\phi/f_a|,
\end{eqnarray}
where we have defined the axion Land\'{e} factor
\begin{equation}
    g_a = \left\langle \frac{\mathbf{S}\cdot\mathbf{J}}{|\mathbf{J}|^2} \right\rangle = \frac{J(J+1) + S(S+1) - L(L+1)}{2J(J+1)}.\label{eq:lande}
\end{equation}
Note that $g_J\neq g_a$.  If we can find states in the atom or molecule where  the values of $g_J/g_a$ are different, then we can use these states as co-magnetometers.  For example, the spin-orbit components $^2P_{1/2}$ and $^2P_{3/2}$ of a $^2P$ electronic state have $g_{J,1/2}=2/3,g_{a,1/2}=-1/3$ and $g_{J,3/2}=4/3,g_{a,3/2}=1/3$, respectively.  Thus, the relative shift due to a magnetic or axion field between these states are not proportional, and they can be distinguished.

Note that not all spin-orbit states have this feature; the $^3P_{0,1,2}$ components of a $^3P$ electronic state all have $g_J=3/2,g_a=1/2$ so comparing the shifts in these states cannot be used to disentangle a magnetic and axion field.  Hyperfine structure, and the fact that $g_e\neq 2$ exactly, make these conclusions not entirely valid, but it means that the $g-$factors differ by $O(10^{-3})$ so their utility as co-magnetometers is suppressed.

Thus, a useful co-magnetometer scheme for the approach under discussion requires states with different relative contributions of electron spin and electron orbital angular momentum to the magnetic moment.  This shows why the internal co-magnetometry scheme for ACME is not immediately useful -- the pairs of states have, to good approximation, the same relative orientation of electron spin and orbital angular momenta.  This is also the case with tuning magnetic interactions in polyatomics with parity doublets~\cite{Anderegg2023CaOHSpin}; these work by changing the average spin projection on the laboratory field, within a single state where the magnetic interactions come almost entirely from the electron spin, and are therefore not immediately useful for axion co-magnetometry.

%%%%%%%%%%%%%%%%%%%%%%%%%%%%%%%%%%%%%%%%%%%%%%%%%%%%%%%%%%%%%%%%%%%%%%%%%%%%%%%%%%%%%%%%%%%%%%%%%%%%%%%%%%%%%%%%%%%%%%%%

 \subsection{Molecular probes}
Polarized diatomic molecules have been used to search for the electron's EDM \cite{Hudson:2011zz,PhysRevLett.119.153001,ACME:2018yjb,Roussy:2022cmp}.
One example is the ACME experiment which sets a bound\footnote{The JILA eEDM experiment~\cite{Roussy:2022cmp} improves the ACME bound by a factor $\sim 2.4$; however, as discussed in the earlier footnote, this experiment uses an ion trap and is therefore not ideal for the measurements discussed here.} on this parameter using the metastable $H$ state in ThO molecule, $|d_e|<1.1\times 10^{-29}$~e~cm \cite{ACME:2018yjb}. This state has $J=1$ 
and enjoys a natural immunity to stray magnetic fields, due to a cancellation between the spin and orbital angular momentum of the valence electrons
which leads to a small net magnetic moment $\mu_H=g_H\mu_B$, with $g_H=0.008$ \cite{2011PhRvA..84c4502V}.   Note, however, that since only the electron spin contributes to axion precession, and the stretched states in the $H,J=1$ manifold have fully-aligned electron spins, this state can still be used to search for the axion gradient.  

The value of $d_e$ is extracted from the change in the precession frequency that is correlated with the molecular axis orientation, given by $\Omega=\textbf{J}_e\cdot \hat{\textbf{n}}$, and the orientation of the effective electric field,  $\omega_{\text{edm}}=d_e E_{\text{eff}}\Omega$, which is reversed every few seconds.  In a later section, we discuss how ACME III (or a similar experiment) could be modified to search for axion forces by searching for spin precession arising from the axion gradient as opposed to the electron EDM.  As mentioned earlier, many of the challenges of electron EDM experiments, such as the need for large electric polarization and the need for heavy species to make use of relativistic enhancements, are not needed; however, since ACME III could make the proposed measurements with minimal modifications, we present the details.  We also discuss simpler dedicated approaches which would not offer electron EDM sensitivity.

For the sake of completeness, we now perform a rough estimate the reach of the experiment in a simple scenario where we source an axion field with a cubic brick of size $D$.
The condition for detecting the axion energy shift is $\Delta E_\phi>\delta \omega$, with $\delta \omega$ the smallest measurable frequency at a given experiment. Let us assume we use a cubic brick of size $D$ made of a dense material with number density of nucleons $n_N$ at a distance $d$ from the molecules. From eq.(\ref{eq:potential}) the energy shift generated by $N\sim n_ND^3$ particles, in the case $D\sim d$, reads:
\begin{equation}
    \Delta E_\phi\sim \frac{g_sg_p^\psi}{8\pi m_\psi}n_N D\left ( \frac{D}{\lambda_\phi }+1\right)e^{-D/\lambda_\phi}\,.
\end{equation}

As an example, axions with wavelength comparable to the other scales in the problem, $2d\sim D\sim \lambda_\phi$, can be detected provided that
\begin{equation}
   g_s g_\psi^p  > \frac{\pi\delta \omega\, m_\psi}{n_N \lambda_\phi}\,,
\end{equation}
which shows how we can gain sensitivity by increasing (decreasing) the parameters $n_N$($\delta \omega$). 

In general, the sensitivity lines at Fig.\ref{fig:axion_electron_forces} and \ref{fig:axion_nucleon_forces} are obtained imposing that the smallest measurable frequency at each of the experiments is equal to the axion-induced energy shift, $\delta\omega=\Delta E_\phi$.
The reach of ACME III can be obtained by rescaling the sensitivity reported at ACME II \cite{ACME:2018yjb}, which was $\delta \omega\sim 480 \mu$rad/s, by the expected improvement factor at ACME III which is around $\sim 30$.\footnote{See Table 1 on the ACME website \href{https://cfp.physics.northwestern.edu/gabrielse-group/acme-electron-edm.html}{https://cfp.physics.northwestern.edu/gabrielse-group/acme-electron-edm.html} for a list of improvements.)}  These improvements give a (shot noise limited) sensitivity at the level of $\delta\omega\sim 15$ $\mu$rad/s and the corresponding sensitivity lines are
%for axion forces induced by the earth gradient and test masses 
shown in Fig.\ref{fig:axion_electron_forces} for axion gradients from the earth and from test masses. We assume lead or tungsten bricks of size $D^3\sim(10~\text{cm})^3$ next to the beam, at a distance of order $O(10)$ cm.   Note these estimates assume that the same statistical and systematic sensitivity for the EDM experiment can be realized for the axion search, which will require a careful experimental analysis given the differences in the protocol.

\section{Experimental setup and background overview}

Molecular beam experiments are well-suited to search for axionic forces on electrons. In this section we first consider using ACME III. We also discuss the experimental set up and the protocol to control systematic effects. Similar strategies are expected for a dedicated axion search using, for example, the Yb atom beam co-magnetometer described in section \ref{sec:Yb-comag}. 
\subsection{ACME III}

An axion gradient generates an additional term to the spin evolution (see Eq. \ref{eq:spin_eq}), with the precessed angle due to the axion contribution given by:
\begin{equation}
    \theta_{\text{axion}}=\int_{0}^{L}\frac{\nabla\phi}{f_\phi}\frac{dx}{v_{mol}}\,, %\label{eq:thetaaxionbeam}
\end{equation}
where $v_{mol}$ is the velocity of the molecules, $x$ is the position, and $L$ is the precession length.  As in EDM searches, the precessed phase can be detected by measuring the population in the spin quadrature states, $S_{x,y}$.

Unlike for EDM searches, the polarizing electric field $E_{\text{lab}}$ is not needed and one could in principle operate with only a weak applied B field. Assuming that the magnetic field is adjusted so that the phase is $\theta_B+\theta_{\text{offset}}\approx \pi/4$, the relevant measurable quantity is given by the asymmetry~\cite{Baron2017ACMELong}:
\begin{align}\label{eq:precession_asymmetry}
    \mathcal{A}=&
    \frac{S_x-S_y}{S_x+S_y}=\mathcal{C}\cos (\theta_B+\theta_\text{axion}+\theta_{\text{offset}})\\&\approx\text{sgn}(B)
% \theta_{\text{edm}}+
\theta_{\text{axion}}=\text{sgn}(B)\left(
%d_e E_{\text{eff}}\Omega+
\frac{\nabla\phi}{f_\phi}\right)\tau_{\text{coh}}\,.
\end{align}
The constant $0\leq \mathcal{C}\leq 1$ is the contrast, which is $\sim 1$ for ACME, and indicates the efficiency in the preparation and detection of the states.   

We discuss two scenarios: one to look for the axion field from a test mass, and one from the earth.  The test mass case is in principle straightforward, as one merely needs to add a moving mass near the ACME beam line.  Let the test mass be movable between positions 1 and 2, sourcing an averaged gradient over the beam path of $-\mathbf{\nabla}\phi_1/f_a$ and $-\mathbf{\nabla}\phi_2/f_a$, respectively.  Considering the test mass position as a binary ``switch'' which can be in state $\mathcal{M}=\pm 1$, analogous to other switches in ACME~\cite{Baron2017ACMELong}, we can write the spin precession angle due to the axion gradient as 
\begin{eqnarray}
    \theta_{\text{axion}} & = & \int_{0}^{L}\frac{\nabla\phi_1+\nabla\phi_2}{2f_\phi}\frac{dx}{v_{mol}} + \mathcal{M}\int_{0}^{L}\frac{\nabla\phi_1 - \nabla\phi_2}{2f_\phi}\frac{dx}{v_{mol}}\,\nonumber\\
    & \equiv & \theta_{0,\text{axion}} + \mathcal{M}\theta_{\mathcal{M}}.
\end{eqnarray}
Note that we have defined a mean, offset spin precession $\theta_{0,\text{axion}}$ which does not depend on the position of the test mass, and a term $\theta_\mathcal{M}$ which changes sign when the test mass is moved.  

The experimental protocol is therefore to add the $\mathcal{M}$ switch where we operate the experiment with the test mass in two positions. 
Moving the mass every few seconds should give sufficient robustness against drifts in other experimental quantities, comparable to other switches in ACME.   This switching protocol is critical for attaining the shot-noise limit on DC quantities as it effectively modulates the DC signal at a higher frequency, thus avoiding effects arising from slow drifts, $1/f$ noise, etc.~\cite{Kirilov2013SNL,Panda2019SNL}.  This should give a spin precession signal which is proportional to the axion field gradient, but possibly other systematic effects.  Note that the masses can be between the preparation and readout stages, so that they won't interfere with the current optical preparation and readout schemes.
 
 One of the effects of most obvious concern is that moving a large mass will change the electromagnetic environment.  Shielding the electric field from the test mass is straightforward: simply use a conducting shield around the molecules, which ACME already has in the form of electric field plates.  The ACME spin precession scheme is fairly robust against electric field drifts by design, as they result in a common-mode offset between the two precessing states, so this is not likely to be a concern.
 
 A greater concern is magnetic offsets.  To get large signals, we would like the test mass to be inside the magnetic shields, as close to the molecules as possible.  Magnetic impurities in the test mass would also correlate with $\mathcal{M}$.  A magnetic field shift of $\sim$1~nanoGauss would give rise to a spin precession signal correlated with $\mathcal{M}$ comparable to the projected statistical sensitivity of ACME III.  Quasi-DC magnetic fields on the order of nanoGauss are challenging to measure, though not impossible; commercially-available optical magnetometers can get to near this sensitivity\footnote{See for example QuSpin, \url{www.quspin.com}}.  However, since these magnetometers work via the interaction of a valence electron on an atom with the magnetic field, they are also in principle sensitive to the axion gradient\footnote{Note that this could be combined with another magnetometer technology not relying on atomic electron spins, such as SQUIDs, as another avenue to search for axion gradients.}.  Thus it is more robust to rely on axion co-magnetometry states, which ACME III is already setting up to use but for a different reason; the $Q^3\Delta_2$ state in ThO~\cite{Wu2020ThOQ} has a magnetic moment of $\sim$2~$\mu_B$, versus $\sim0.01$~$\mu_B$ for the $H^3\Delta_1$ state, and since the magnetic moment arises mostly from orbital angular momentum, the states $H$ and $Q$ represent an axion co-magnetometry state pair (see appendix \ref{sec:Appendix_co-mag} for details). 
Thus, one can measure the spin precession dependence on $\mathcal{M}$ in both the $H$ and $Q$ states; since they have different relative magnetic and axion gradient sensitivity, the relative contributions of these two effects will be different in these two states, thus enabling their disentanglement.  Note that for cases where it is technically feasible, the mass could be periodically rotated or re-oriented to change the direction of the residual fields for further rejection of systematic effects.

A related concern is magnetic Johnson noise (MJN)~\cite{Lamoreaux1999MJN,Munger2005MJN}, which arises due to thermal fluctuation currents in a conductor at finite temperature.  This will not necessarily add systematically offset spin precession, but it will result in magnetic field noise which could reduce the contrast and statistical sensitivity of the measurement.  The calculation of the effect would depend on the specific geometry and material chosen for the moving masses, but we can make some estimates.  For a conductor with resistivity $\rho$ at temperature $T$ having thickness $\sim t$ a distance of $\sim D$ away, the magnetic field power spectral density at the molecules is given approximately by~\cite{Vutha2011Thesis}
\begin{equation}
    \widetilde{\mathcal{B}}(f) \sim \frac{\mu_0}{4\pi}\left[\frac{8 t k_B T}{3\rho D^2}\right]^{1/2}\sim \frac{1~\textrm{pG}/\sqrt{\textrm{Hz}}}{\sqrt{\rho/(1~\Omega\cdot\textrm{m})}},
\end{equation}
where $k_B$ is Boltzmann's constant, and for the rightmost term we have assumed $D\sim$~50~mm, $t\sim$~100~mm, and $T=$~300~K (though the mass could be cooled if needed).

Tungsten~\cite{CRC2022} would be a natural choice for a test mass given its very high mass density of 19.3~g/cm$^3$, and its  resistivity of 5.44$\times10^{-8}~\Omega\cdot$m would give rise to MJN on the order of $\sim$5~nG/$\surd$Hz.  This would give rise to magnetic spin precession noise on the order of $\sim2\pi\times\mu$Hz/$\surd$Hz for the ThO H state, and around a hundred-fold larger for the ThO Q state, which might sound problematic; however, as it is still smaller than other more dominant forms of noise, it will not be a limitation as it will average away faster than other noise sources.

One such source is the velocity dispersion $\Delta v$ in the molecular beam, which also is not a fundamental limitation as it averages away~\cite{Baron2017ACMELong}, though it can add excess noise~\cite{ACME:2018yjb}.  The phase noise $\Delta\phi$ from magnetic spin precession in magnetic field noise $\Delta\B$ and precession time noise $\Delta\tau$, is given by
\begin{equation}
    \Delta\phi=\tau\Delta\B+\B\Delta\tau = \tau\left(\Delta\B + \B\Delta v/v\right).
\end{equation}
We have used the fact that $|\Delta \B|\ll |\B|$ as the applied (and residual) magnetic fields are larger than the MJN, and that the precession is over $L$ so $L=v\tau$ and therefore $\Delta\tau/\tau=\Delta v/v$.  Since the ACME beam has $\Delta v/v\sim0.1$~\cite{Hutzler2011ThO} due to velocity dispersion within a single shot, and has shot-to-shot changes in the mean velocity of $\Delta v/v\sim10^{-3}$~\cite{ACME:2018yjb}, the $\Delta \B$ component should not be a major limitation to statistical sensitivity.  However, should MJN (or the cost of tungsten) ultimately be a limiting factor, there are materials such as zirconia or leaded glass with over 10 orders of magnitude larger resistivity yet only a factor of 4 to 5 less density.

\label{sec:Exp_scheme}

Now we describe a protocol to measure an axion gradient from the Earth. In this case there is no moving mass, and therefore no $\mathcal{M}$ switch.  This introduces a challenge, as we no longer have a way to change the axion field, and are therefore potentially susceptible to the many DC drifts and offsets in EDM-style spin precession experiments~\cite{Baron2017ACMELong}.  However, we discuss two methods which can help mitigate this.

We can continue to the use the $H,Q$ co-magnetometer pair to distinguish between a constant background magnetic field and the background axion field.  A greater challenge is absolute phase offsets, for example arising from the fact that the state initialization and readout stages will always have some finite, drifting offset set by the polarization of lasers with different beam paths.  To address this, we propose to use the fact that the velocity dispersion in the molecular beam~\cite{Hutzler2011ThO,Hutzler2012Review,Hutzler2014Thesis,Ho2020YbFImprove} results in an accumulated phase angle which is time-dependent relative to the time after the production of the molecular beam pulse at $t=0$~\cite{Baron2017ACMELong}:
\begin{equation}
    \theta_{\text{axion}}(t)=\int_{0}^{L}\frac{\nabla\phi}{f_\phi}\frac{dx}{v_{mol}(t)}\,. %\label{eq:thetaaxionbeam}
\end{equation}
This arises from the fact that slower molecules take longer to get to the spin precession region, and when they arrive they spend more time precessing and therefore accumulate more phase.  Because the ACME spin precession readout protocol involves rapid, time-resolved readout to normalize against molecular beam yield fluctuations~\cite{Kirilov2013SNL}, this also gives the ability to resolve this time-dependence on a single shot~\cite{Hutzler2014Thesis}.  This has the advantage of offering significant robustness against preparation and measurement phase errors, which will be constant, and physical spin precession phases, which will have a time-dependence.  Note that inferring the spin precession from the asymmetry time-dependence also provides robustness against sources of offsets such as light shifts from the lasers themselves, which will not accumulate over the entire spin precession period (and which can be probed by varying laser parameters.)  A careful experimental characterization of this protocol would be needed to determine whether it could be used to reach the same frequency sensitivity as the EDM protocol

Therefore, the proposed protocol to measure the axion gradient from the earth is the following.  Note that this protocol is in addition to the established ACME methods for precision measurement of molecular phases, see for example~\cite{Baron2014ACMEI,Baron2017ACMELong,ACME:2018yjb}:

\begin{itemize}
    \item Switch between the $H$-state and $Q$-state in periods of around 1~second.  This will enable robust co-magnetometry, in particular the measurement of background magnetic fields.
    \item Measure the time dependence of the asymmetry, in particular its slope: $\frac{\partial \mathcal{A}}{\partial t}$. 
    \item Compare the asymmetry slope for the $Q$ and $H$ states.  The axion field should cause a component of $\frac{\partial \mathcal{A}}{\partial t}$ which changes between $Q$ and $H$, but otherwise does not change.
\end{itemize}

An important observation is that the ACME quantization axis, which is set by the electric field, is horizontal and therefore has vanishing sensitivity to the Earth axion gradients.  There are two potential approaches to address this.  One could rotate the electric field plates so that the applied field aligns with gravity.  This would be highly non-trivial, not only because the plate mounting would have to be re-engineered, but all the laser paths would need to be re-engineered as there are some lasers which must go through the plates and some which cannot.  Another option would be to operate without an electric field at all, and use a weak vertical magnetic field to set the quantization axis.  This would require modifying the state preparation and readout protocols, but possibly in a way which would not require major redesign of the apparatus.

\subsection{Ultracold atoms and molecules}

There are proposals to use ultracold molecules to search for the electron EDM surpassing the current bounds by several orders of magnitude \cite{Kozyryev:2017cwq,Fleig2021Ag,Fitch2021YbF}.  The shot noise-limited uncertainty of the frequency in a measurement  is given by $\delta\omega=\frac{1}{\tau_c\sqrt{N}}$, with $N$ the number of measured molecules.
Assuming $N=10^6$ molecules and coherence times around 10-100 seconds, just one measurement would be equivalent to the expected sensitivity at ACME III. With a preparation/detection efficiency around $O(10)$\% it is expected to have sensitivities of the order $\delta\omega \sim 1-10$ nrad/s by operating for around $10^7$ seconds. These numbers suggest the ultracold molecule proposal as being very compelling to test axion forces on electrons. 

To achieve full sensitivity it would be important to have a co-magnetometer species that reduces the impact of systematic effects. Co-trapped species~\cite{Gregory2021,Jurgilas2021}, with different origins of the magnetic moment, would allow the distinction between stray magnetic fields and the sourced axion gradient.

It may also be interesting to consider for axion force searches the $^{171}$Yb optical trap in \cite{PhysRevLett.129.083001}, which used the ground state, $^1S_0$, to look for a permanent atomic EDM.  In that work the authors show that the coherence time exceeds $\tau_c\sim O(100)$ seconds, implying that if a number large number of atoms, $N\sim O(10^6)$, can be trapped the sensitivity to monopole-dipole forces on nucleons may extend beyond astrophysical constraints (see Fig.\ref{fig:axion_nucleon_forces}). This scheme may also require the presence of a co-magnetometer species.

%%%%%%%%%%%%%%%%%%%%%%%%%%%%%%%%%%%%%%%%%%%%%%%%%%%%%%%

\subsection{Atomic beam experiments}
\label{sec:Yb-comag}
 
Atom beam experiments are also excellent candidates to look for axion forces. Since there is no ``molecular enhancement'' of the axion signal, unlike EDM searches, it is attractive to use atoms as their simpler structure generally leads to more intense beams, more efficient optical control, easier laser cooling for beam brightening, etc.  Note that an experiment built for the purpose of searching for axion gradients would not need an electric field at all, further simplifying the experimental requirements.

An interesting possibility is an experiment using $^{171}$Yb, which can be used to make intense beams~\cite{Wright2022,Hutzler2012Review}, and has two valence electrons in the $6s$ orbital $(L=S=0)$ giving rise to a $^1S_0$ ground electronic state, and nuclear spin $I=1/2$. The $^3P_2$ $(L=S=1)$ excited state is relatively long-lived with a life-time around $\tau\sim 10$ s~
\cite{Porsev2004AELifetimes}.

\begin{figure*}[t]
	\centering
	\includegraphics[width=0.75\textwidth]{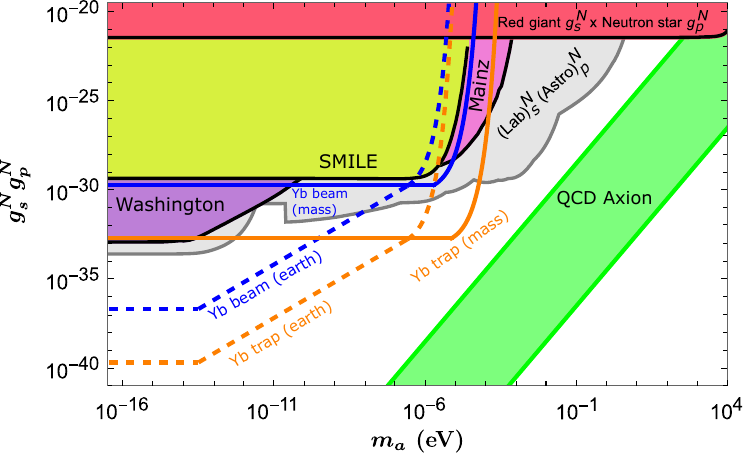}
		\caption{Axion-mediated monopole-dipole forces on nucleons at spin precession experiments searching for axion gradients source by a test mass or the earth. The blue lines stand for expected sensitivity of a dedicated spin precession experiment using Yb co-magnetometer beams assuming that systematic effects are kept under control. Finally, in orange, we have the expected reach with an experiment using cold trapped Yb atoms, assuming the existence of co-magnetometer states. Bounds adapted from \cite{OHare:2020wah}}

	\label{fig:axion_nucleon_forces}
\end{figure*}

An experiment could use the ground state $^1S_0$, which is only sensitive to the axion gradient through the nucleon spin, and the excited state $^3P_2$, which has a very different magnetic moment and can therefore be used as a co-magnetometer.  The Hamiltonians for these states are:
\begin{align}
    &H_{^1\!S_0}=-\mathbf{\mu_N}\cdot\mathbf{B}+c_N\frac{\mathbf{\nabla}\phi}{f_\phi}\cdot\mathbf{I}\,,\\&
    H_{^3\!P_2}=-(\mathbf{\mu_N}+M g_P\mathbf{\mu_B})\cdot\mathbf{B}+\frac{\mathbf{\nabla}\phi}{f_\phi}\cdot(c_N\mathbf{I}+c_e\mathbf{S})\,,
\end{align}
where $\mu_{N(B)}$ is the nuclear (Bohr) magneton, $M$ is the projection of the total angular momentum onto the quantisation axis, and $\mathbf{I}$ ($\mathbf{S}$) is the nucleon  (electron) spin. The coefficients $c_e, c_N$ reflect the fact that in principle the axion coupling to electrons and nucleons may differ. These states have different origins for their magnetic moment and can be used as a co-magnetometer by comparing how the precession frequency changes with the B field and axion gradient orientation, as discussed in a previous section. In the notation described in Sec.\ref{sec:axion_comagnetometry}, this corresponds to $g_{a,0}=0$ and $g_{F,0}\sim 10^{-3}$ \cite{Olschewski1972} for the $^1S_0$ state, where $F=I+J$ is the total angular momentum, and $g_{F,2}
\approx 3/2$, $g_{a,2}=1/2$ for the $^3P_2$.  Since the ground state is only sensitive to the axion coupling to nucleons, this experiment would be sensitive to both coupling to electrons and nucleons. See Figures \ref{fig:axion_electron_forces} and \ref{fig:axion_nucleon_forces}.  Note that in the event of a positive signal, it would be critical to perform the measurements in different states or species with additional different relative sensitivities to magnetic fields, electron couplings, and nuclear couplings, in order to conclusively disentangle these effects.

%%%%%%%%%

The shot noise-limited sensitivity using
$\tau = 5 - 10$~ms, $\dot{N}=10^{10}-10^{11}$ atoms/s, and $T_{int}=10^6 - 10^7$~s, we get an expected sensitivity in the range $\delta\omega\sim 10^{-7}$ Hz to $10^{-6}$ Hz. In Fig.\ref{fig:axion_electron_forces} we show the expected reach assuming $\delta\omega\sim 10^{-7}$ Hz.

Alternatively one could use an indium or thallium beam using the $^2P_{1/2}$ and $^2P_{3/2}$ spin orbit components of the ground electronic state. As discussed earlier, these two states can be used as as axion co-magnetometers since their $g$-factors have different contributions from electron spin and orbital angular momenta 
giving: $g_{J,1/2}=2/3,g_{a,1/2}=-1/3$ and $g_{J,3/2}=4/3,g_{a,3/2}=1/3$. Whether they offer any advantage over Yb depends largely on experimental considerations, such as beam fluxes, laser wavelengths, detection strategies, etc. 

 \section{Advantages of beam experiments}

ACME III will measure the spin precession at the level  of approximately 10 microrad/s. This corresponds, in terms of a Bohr magneton frequency, to a magnetic field of around 10 aT. It is interesting to compare this result to similar searches using co-magnetometers (see Fig.\ref{fig:axion_electron_forces}). 
In the SMILE experiment \cite{Lee:2018vaq} an alkali-noble co-magnetometer is employed. These detectors are currently one of the most sensitive magnetometers, able to measure magnetic fields at the level of $O(1-10)\text{fT}/\sqrt{\text{Hz}}$.

The spin projection noise, given by
\begin{equation}
    \delta B=\frac{1}{\mu_B}\frac{1}{\sqrt{T_2 t_{int} N}}\,,
\end{equation}
indicates that, in principle, this kind of co-magnetometer would surpass the ACME III expected sensitivity  using the values: $T_2\sim 10^{-3}$~s, $N\sim 10^{13}$ and $t_{int}\sim 10^5$~s. 

However, the co-magnetometer sensitivity  is limited by optical noise of the probe laser which, at low frequences below $\sim 0.5$ Hz, goes as 1/f (see section 4.2.6 in \cite{Lee:2019ixj} for details).
At frequencies around 0.1 Hz (given by the moving mass period) dominates and lies around $\sim 10 $ fT/$\sqrt{\text{Hz}}$. For an integration time $t_{int}\sim 10^5$ s this leads to an uncertainty  $\delta B\sim $ 70 aT for the effective magnetic field \cite{Lee:2018vaq}, i.e. the axion gradient, a bit less than an order of magnitude larger than the shot-noise limited sensitivity expected at ACME III. 

This explains why in terms of reach,  the use of EDM experiments like ACME III, with all systematics
eliminated, is expected to improve the current co-magnetometer results \cite{Lee:2018vaq} by around an order of magnitude if moving test-masses are employed. Additionally,   thanks to the time-dependence of the asymmetry, one can use the \textit{imperfect behaviour} of the molecular beam in terms of velocity dispersion to measure the Earth gradient as  discussed above. This fact will further improve the reach by several orders of magnitude, in particular for light axions. For axions with mass $m_a\lesssim 10^{-10}$ eV, regions of the parameter space beyond the constrained by astrophysics may be tested.

\section{Conclusion}

One of the generic ways in which new physics can interact with the Standard Model is through the spin of standard model particles. Spin precession experiments are thus well placed to search for a variety of such effects, ranging from time varying effects caused by dark matter to new fields sourced by terrestrial and laboratory sources. For the latter class of experiments, the signal is fundamentally a DC signal. A variety of low frequency systematic effects must be overcome in order to see this signal. Interestingly, there are spin precession experiments that are adept at managing such low frequency systematic effects -- namely, experiments that are aimed at measuring the permanent electric dipole moment of electrons and nucleons. In this paper, we have highlighted the opportunities that exist in using the well developed technology of experiments such as ACME III, as well as motivated to build dedicated experiments using atomic beams, and laser-cooled atoms or molecules to search for spin precession induced by test masses and the earth in the laboratory.

\section*{Acknowledgments}
\noindent PA is supported by the STFC under Grant No. ST/T000864/1. 
N.R.H. is supported by U.S.~National Science Foundation (NSF) CAREER Award (PHY-1847550), the Heising-Simons Foundation (2022-3361), the Gordon and Betty Moore Foundation (GBMF7947), and the Alfred P. Sloan Foundation (G-2019-12502).
D.~E.~K and S.R.~are supported in part by the NS under Grant No.~PHY-1818899.  
This work was supported by the U.S.~Department of Energy (DOE), Office of Science, National Quantum Information Science Research Centers, Superconducting Quantum Materials and Systems Center (SQMS) under contract No.~DE-AC02-07CH11359. S.R.~is also supported by the DOE under a QuantISED grant for MAGIS, and the Simons Investigator Award No.~827042. This article is based upon work from COST Action COSMIC WISPers CA21106, 
supported by COST (European Cooperation in Science and Technology).
NH and MR thank the Perimeter Institute and the organisers of the \textit{School on Table-Top Experiments for Fundamental Physics}, where this work was initiated, for providing a friendly and exciting atmosphere.

\appendix

\section{Axion co-magnetometry in ThO}\label{sec:Appendix_co-mag}

We will always have stray magnetic fields, $\mathbf{B}_{str}$, and if $\mathbf{B}_{str}$ has a component along the radially oriented Earth axion gradient then one could erroneously identify this as an axion signal. Crucially, co-magnetometry techniques allows us to suppress the leading order contribution of the stray magnetic fields to the spin precession frequency. 

The precession frequencies of the H- and Q-state of ThO taking into account the axion gradient and stray magnetic fields are given by:
\begin{equation}
    \omega_{H,Q}=-g_{H,Q}\mathbf{\mu}_B\cdot(\textbf{B}_{str})-\mathbf{S}\cdot\frac{\nabla \phi}{f_\phi}\,.
\end{equation}
In practice, ACME III will use the H-state and Q-state of the ThO molecule, where $g_Q\approx 2$ and $g_H=8.8 \times 10^{-3}$~\cite{Wu2020ThOQ}.  Since the Q-state has nearly zero spin projection over the internuclear axis~\cite{Wu2020ThOQ}, the axionic g-factor of the Q-state is also expected to be small $g_a^Q\ll 1$, only due to small corrections (e.g. spin-orbit mixing with other electronic states~\cite{Petrov2014ThOZeeman}).  Thus, the procedure is to measure spin-precession in the $Q$-state, obtaining $\omega_Q\approx -g_Q\mu_BB_{str}$

Since the gyromagnetic ratios $g_H$ and $g_Q$ only appear multiplicatively in the signal, we do not require them to be measured to the same accuracy as other quantities.  Specifically, the axion signal and its uncertainty are given by
\begin{eqnarray}
    \mathbf{S}\cdot\nabla \phi/f_\phi & = & -\omega_H -\mu_H B
     = -\omega_H -\mu_R \omega_Q \\
    \delta(\mathbf{S}\cdot\nabla \phi/f_\phi)^2 & = & (\delta\omega_H)^2 + (\delta \omega_Q)^2\mu_R^2 + (\delta \mu_R)^2\mu_Q^2B^2
\end{eqnarray}
where we have defined the ratio $\mu_R=\mu_H/\mu_Q\approx 10^{-2}$.
Assuming that $\delta\omega_H \approx \delta\omega_Q$, we can see that the second term will be negligible.  The uncertainty in the ratio is $(\delta\mu_R/\mu_R)^2\approx(\delta\mu_{H,Q}/\mu_{H,Q})^2$, assuming that the fractional uncertainty on the individual moments is comparable.  Therefore, the uncertainty on the moments will not dominate as long as
\begin{equation}
    \delta\omega \gtrsim \mu_H B (\delta\mu_{H,Q}/\mu_{H,Q}).
\end{equation}
Therefore, if $B\lesssim$0.1 microgauss and $(\delta\mu_{H,Q}/\mu_{H,Q})\lesssim 10^{-3}$, this should not be a limitation, and both values could potentially be improved in future measurements.

In the case where a magnetic field is purposefully applied, the precession frequencies of the H- and Q-state of ThO taking into account the applied magnetic field, the axion gradient, and stray magnetic fields are given by:

\begin{equation}
    \omega_{H,Q}=-g_{H,Q}\mathbf{\mu}_B\cdot(\textbf{B}_{lab}+\tilde{\textbf{B}}_{str})-\mathbf{S}\cdot\frac{\nabla \phi}{f_\phi}\,.
\end{equation}

The stray magnetic field $\tilde{\textbf{B}}_{str}$ effect can be canceled by taking combinations of the type:
\begin{equation}\label{eq:comb_freqs}    \frac{\omega_H^{\text{up}}}{\omega_Q^{\text{up}}}-\frac{\omega_H^{\text{down}}}{\omega_Q^{\text{down}}}=\frac{2\omega_{axion}}{g_Q\mu_BB_{lab}}(g_a^H-\frac{g_H}{g_Q}g_a^Q)+O(\tilde {B}_{str}^3/B_{lab}^3)\,.
\end{equation}
where ``up/down" stand for the orientation of the spins relative to the vertical direction (quantisation axis, defined by $\mathbf{B}_{lab}$), and $g_a^{H,Q}$ are the \textit{axionic} g-factors (see text). This combination of frequencies for both quantum states is only non-zero if there is an axion gradient and  $g_H\neq g_Q$.  
The error budget from stray magnetic fields is $\mathcal{O}(\tilde {B}_{str}^3/B_{lab}^3)$, with linear and quadratic effects in $\tilde{B}_{str}$ cancelling due to co-magnetometry. 
In ACME I and II, the applied magnetic field was around $B_{lab}=10^{-3}$ Gauss. This implies that as long as stray magnetic fields at the interaction region are shielded at the level of $10^{-11}$ T=0.1 microGauss, then the systematic effect induced by stray magnetic fields will be equivalent to magnetic fields   $\ll 10^{-18}$ T, which should suffice to achieve the expected sensitivity.

%\bibliographystyle{utphys}
%\bibliography{newrefs_axion.bib}

\providecommand{\href}[2]{#2}\begingroup\raggedright\endgroup

\end{document}